\documentclass[aps,prc,superscriptaddress,showpacs,nofootinbib,floatfix,twocolumn]{revtex4}

\usepackage{doi,graphicx,amssymb,amsmath,dcolumn}

\newcommand{\etal}{\textit{et al. }}

\begin{document}

\title{Disordered nuclear pasta, magnetic field decay, and crust cooling in neutron stars}
\author{C. J. Horowitz}\email{horowit@indiana.edu}
\affiliation{Department of Physics and Center for the Exploration of Energy and Matter, Indiana University, Bloomington, IN 47405, USA}
\author{D. K. Berry}
\affiliation{University Information Technology Services, Indiana University, Bloomington, IN 47408, USA}

\author{C. M. Briggs}
\author{M. E. Caplan}
\affiliation{Department of Physics and Center for the Exploration of Energy and Matter, Indiana University, Bloomington, IN 47405, USA}
\author{A. Cumming}
\affiliation{Department of Physics, McGill University, 3600 rue University, Montreal QC, H3A 2T8 Canada}
\author{A. S. Schneider}
\affiliation{Department of Physics and Center for the Exploration of Energy and Matter, Indiana University, Bloomington, IN 47405, USA}
\date{\today}
\begin{abstract}
Nuclear pasta, with non-spherical shapes, is expected near the base of the crust in neutron stars.  Large scale molecular dynamics simulations of pasta show long lived topological defects that could increase electron scattering and reduce both the thermal and electrical conductivities.  We model a possible low conductivity pasta layer by increasing an impurity parameter $Q_{imp}$.  Predictions of light curves for the low mass X-ray binary MXB 1659-29, assuming a large $Q_{imp}$, find continued late time cooling that is consistent with Chandra observations.  The electrical and thermal conductivities are likely related.  Therefore observations of late time crust cooling can provide insight on the electrical conductivity and the possible decay of neutron star magnetic fields (assuming these are supported by currents in the crust). 
\end{abstract}


\maketitle

Complex nuclear pasta phases are expected near the base of the crust in neutron stars.  Nuclear matter with many non-spherical shapes  is possible because of competition between short range nuclear attraction and long range Coulomb repulsion \cite{Pethick19987} that gives rise to Coulomb frustration, see for example \cite{PhysRevLett.41.1623,PhysRevC.69.045804,PhysRevC.83.035803}.  Recently Pons \etal \cite{pons2013highly} suggested that this pasta may have a high electrical resistivity that could lead to magnetic field decay in neutron stars.   When the field decayed, the star would stop spinning down from electromagnetic radiation.  This could explain why few X-ray pulsars are found with spin periods longer than 12 seconds \cite{pons2013highly}.  

Nuclear pasta is a charge neutral system of neutrons, protons and electrons.  Because of Pauli blocking, the degenerate electrons are expected to have a relatively long mean free path.  Therefore, electron transport likely dominates the system's electrical conductivity, thermal conductivity, and shear viscosity \cite{PhysRevC.78.035806}.   This suggests a relationship between these three transport coefficients.  Indeed for conventional metals, the relation between electrical and thermal conductivities is known as the Wiedemann-Franz law \cite{Wiedemann1853}.   However, the transport of superfluid neutrons may make a subdominant contribution to the thermal conductivity (but not the electrical conductivity) \cite{Aguilera2009}.

The thermal conductivity of neutron star crust can be probed with X-ray observations of crust cooling after extended periods of accretion, see for example \cite{Brown2009}.  The rapid crust cooling observed over time periods of a year or less suggest that at least the outer crust has a high thermal conductivity and is likely to be crystalline and not amorphous.  These observations of crust cooling are particularly powerful because the cooling over different time scales is sensitive to the thermal conductivity at different depths in the crust.  In particular, crust cooling, three to ten years after accretion stops, may be sensitive to the thermal conductivity at densities near $10^{14}$ g/cm$^3$ where nuclear pasta is expected.  

Does nuclear pasta, in fact, have small electrical and thermal conductivities?  Pons \etal parameterize the electrical conductivity with an impurity parameter $Q_{imp}$.  Strictly speaking, this assumes a uniform crystal lattice of ions with a distribution of charge for each ion that is characterized by $Q_{imp}$.  Pasta may not involve a simple crystal lattice.  Furthermore, nucleons may be free to move into and out of particular pieces of pasta to equilibrate their compositions.  Therefore, this impurity parameter formalism is not directly applicable to nuclear pasta.  Nevertheless, it may still provide a simple way to parameterize possible effects of disorder on the electrical and thermal conductivities.

What could lead to disorder in nuclear pasta and decrease its conductivities?  We present molecular dynamics simulations of nuclear pasta and explore the formation of topological defects.   Similar defects may have been observed by Alcain et al. \cite{Alcain2014}.  To determine the effect of these defects on the thermal conductivity we make a simple estimate of an effective impurity parameter $Q_{imp}$ based on our MD results.  Finally, we demonstrate the sensitivity of crust cooling X-ray light curves to this $Q_{imp}$.  



We use molecular dynamics simulations to explore the formation of topological defects.  There are a number of quantum calculations of nuclear pasta structure based on density functionals, see for example \cite{2009arXiv0904.4714S},\cite{PhysRevC.87.055805},\cite{Schutrumpf:2014vqa}.  However, these have limited simulation volumes that may be too small to recognize defects.  Instead we present classical molecular dynamics calculations using a simple model \cite{2011arXiv1109.5095H} that allows much larger simulation volumes.  Our model has been used previously to describe neutrino scattering \cite{PhysRevC.69.045804,PhysRevC.70.065806}, the dynamical response function \cite{PhysRevC.72.035801}, and other transport properties \cite{PhysRevC.78.035806}.  The formation of complex pasta phases was explored in ref. \cite{PhysRevC.88.065807}.  In the model nucleons interact via a two-body potential,
\begin{equation}
 V(r)=a {\rm e}^{-r^2/\Lambda}+[b\pm c]{\rm e}^{-r^2/2\Lambda}+\frac{e_ie_j}{r}{\rm e}^{-r/\lambda}\, .
\end{equation}
Here the plus sign is for the interaction between two protons or two neutrons while the minus sign describes the more attractive interaction between a proton and a neutron.
In the equations above, $e_i$ is the charge of the $i^{th}$ nucleon, $r$ is the distance between the two nucleons and $a$, $b$, $c$ and $\Lambda$ are constants adjusted to approximately reproduce some bulk properties of pure neutron matter and symmetric nuclear matter as well as the binding energies of selected nuclei \cite{PhysRevC.69.045804}.  The Coulomb interaction is screened by the slightly polarizable electron gas.  For simplicity we use a fixed screening length $\lambda=10$ fm that was employed in previous works.

We simulate at a temperature $T=1$ MeV that has been previously studied.  Neutron star crust has a much lower temperature.  However dynamics at 1 MeV may be much faster and accessible to direct MD simulation, while defect formation at lower temperatures may be too slow to directly simulate.  We also use a proton fraction $Y_p=0.4$ because this has been used in previous work.  Neutron star crust in beta equilibrium has a lower proton fraction and this will be studied in future work.   Finally, we simulate at a baryon density of $n=0.05$ fm$^{-3}$ or $1\times 10^{14}$ g/cm$^3$.  This is about 1/3 of nuclear saturation density, and is a typical density where nuclear pasta is expected.  All of our simulations use a MD time step of 2 fm/c.

\begin{figure}
\centering
\includegraphics[width=0.5\textwidth]{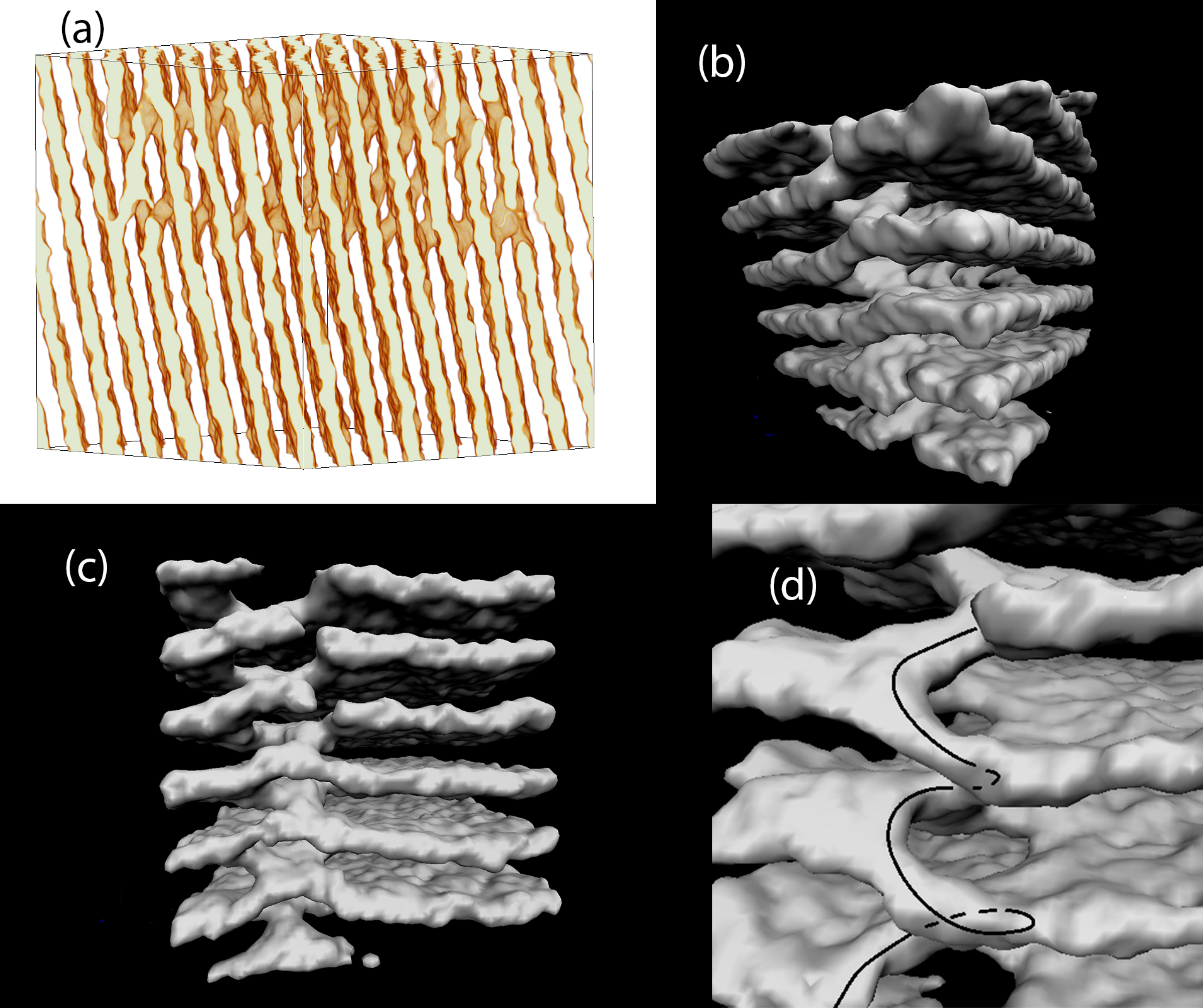}
\caption{\label{fig.messy} (Color online) (a) Proton density isosurface for a configuration of 409600 nucleons from an MD simulation that started from a random configuration.  (b) Proton density isosurface for a 50,000 nucleons MD simulation.  Flat lasagne plates to the left and right are separated by a plane of topological defects that extend from the front left to the back right and from the top to the bottom. (c) Proton density isosurface for a 75,000 nucleon MD simulation.  A single screw defect near the left foreground extends vertically throughout the simulation volume.  (d) Closeup of a single defect showing its helical nature.}
\end{figure}

Figure \ref{fig.messy} (a) shows the final configuration of 409,600 nucleons after evolution for $1.4\times 10^7$ fm/c.  The starting point was a uniform random configuration.  The result is flat plates that are connected by a number of topological defects.  We will explore these defects below.  The defects form rather quickly and then are stable for the full remaining simulation time.

Next we present the final configuration of a simulation with 50,000 nucleons in Fig. \ref{fig.messy} (b).  Starting from a uniform random initial configuration the nucleons quickly formed an array of screw like defects that formed a plane near the center of the simulation extending from left front to right back and top to bottom.  To the left and right of this array of defects are normal flat plates that are displaced by half a lattice spacing in crossing the array.  

Fig. \ref{fig.messy} (c) shows a simulation with 75,000 nucleons that started from a uniform random configuration and was evolved for about $1.6\times 10^7$ fm/c.  This simulation shows flat plates that are connected by a single screw like topological defect near the right foreground.  This defect extends vertically throughout the simulation volume.  It formed quickly and then remained stable as it slowly drifted in the simulation volume.  Finally, Fig. \ref{fig.messy} (d) shows a closeup of a single defect that shows its helical nature.  This appears simillar to screw defects in liquid crystals, see for example Ref. \cite{Chaikin} Fig. 9.2.5.

Additionally, we perform simulations varying $T$, and $Y_{P}$, to study the stability of these defects. These 51200 nucleon simulations were equilibrated from random initial conditions for 500,000 fm/c. We find that 10\% variation of temperature (simulations with $T$=0.9 and $T$=1.1 MeV) still produces defects similar to those shown in Fig \ref{fig.messy} (c) and (d), as do 10\% variations in proton fraction (simulations with $Y_{p}$ = 0.35, 0.45).   Note that there were a few simulations with very different parameter values that did not appear to form defects.


In summary, many of our MD simulations formed topological defects either as a single column, Fig. \ref{fig.messy} (c), as a two dimensional array, Fig. \ref{fig.messy} (b), or at multiple sites in the simulation volume as in Fig. \ref{fig.messy} (a).  These defects tend to form quickly, when simulations are started from random positions, and then persist for long simulation times.


We now make a rough estimate, based on Fig. \ref{fig.messy} (b), of how these defects could act as effective impurities and reduce the thermal and electrical conductivities.  The charge distribution in the central region of Fig. \ref{fig.messy} (b) differs greatly from the normal pasta configuration (that is present to the left and right).  In the limit of very large impurities, $Q_{imp}=<Z>^2$.  Here $<Z> \approx 20$ is the mean effective charge of a cluster (piece of pasta), see below.  Our simple model for an effective $Q_{imp}$, that might describe electron-pasta scattering from the configuration in Fig. \ref{fig.messy} (b), is 
\begin{equation}
Q_{imp} \approx f_D <Z>^2 \approx 40\, .
\label{eq.qimp}
\end{equation}
Here $f_D$, of order $0.1$, is the fraction of the volume of the total system that is occupied by defects.  

The effective charge, that an electron can scatter from, is limited at low momentum transfers $q$ by screening and at high momentum transfers by the elastic form factor of a single piece of pasta.  As a result $<Z>$ is the maximum (as a function of $q$) of the angle averaged proton static structure factor $S_p(q)$.  This has been calculated in ref. \cite{PhysRevC.78.035806}, at a density 0.05 fm$^{-3}$, $<Z>=S_p(q)\approx 20$.   Note that we expect the conductivity, for nearly pure pasta, to increase as the temperature decreases.  Therefore we do not use these calculations of $S_p(q)$, done at $T=1$ MeV, directly to calculate the thermal conductivity at lower temperatures.  

We emphasize that the rough estimate $Q_{imp}\approx 40$ in Eq. \ref{eq.qimp} is uncertain.  The MD simulation results likely depend on the size of the simulation, the initial conditions, and on various thermodynamic parameters.  Perhaps the most difficult problem, and a large source of uncertainty, is to determine (from first principles) the actual density of defects.  Nevertheless, we are not aware of any other estimates of $Q_{imp}$ for pasta.  

Therefore, we explore some implications if $Q_{imp}$ is indeed of order 40.  First, this value is roughly comparable to $Q_{imp}$=100 considered by Pons et al.~\cite{pons2013highly} for the electrical conductivity of pasta and the evolution of magnetic fields.  Presumably, following Pons et al., a value of $Q_{imp}\approx 40$ would likely lead to decay of magnetic fields, in of order $10^6$ years (if these fields are supported by currents in the crust).

Second, we explore the implications of $Q_{imp}\approx 40$ for the thermal conductivity of pasta and for crust cooling after extended periods of accretion.  We model the effect of a pasta layer with low thermal conductivity on the cooling curve by following the thermal relaxation of the crust. We explore if the disordered pasta layer results in late time cooling at times of thousands of days into quiescence.  We use a numerical method similar to Brown et al. \cite{Brown2009} with the same input microphysics for the crust.   We include the low conductivity layer by increasing the impurity parameter $Q_{\rm imp}$ at high density $\rho_{\rm pasta}<\rho<\rho_{\rm core}$ where $\rho_{\rm core}$ is the crust-core transition density. Note that this implicitly assumes a particular temperature and density dependence for the thermal conductivity of the pasta, with thermal conductivity increasing with temperature (see discussion in Page \& Reddy \cite{Page2012}\cite{Page2013}).

We compare our models with the temperature measurements for MXB~1659-29 presented by Cackett et al.~\cite{Cackett2008}. They analyzed a Chandra observation of MXB~1659-29 taken almost 11 years into quiescence, 4 years after the previous Chandra observation. They found that the count rate had decreased by a factor of 3 in that time, with a clear change in the shape of the spectrum.  As they discuss, two explanations that are consistent with the observed change in the spectrum are either that the absorption column increased between the two observations, while the neutron star held a fixed temperature, or the neutron star thermal component cooled significantly, and a power law component developed that dominates the spectrum. Cackett et al.~\cite{Cackett2013} argued that a changing $N_H$ was preferred by crust cooling models, which predict a constant temperature at times $\gtrsim 1500$ days when the neutron star crust has come into thermal equilibrium with the core (\cite{Cackett2008}; \cite{Brown2009}). We show here, however, that a low conductivity layer at the base of the crust does result in late time cooling that can match the latest observation of MXB~1659-29.

\begin{figure}[ht]
\centering
\includegraphics[width=0.5\textwidth]{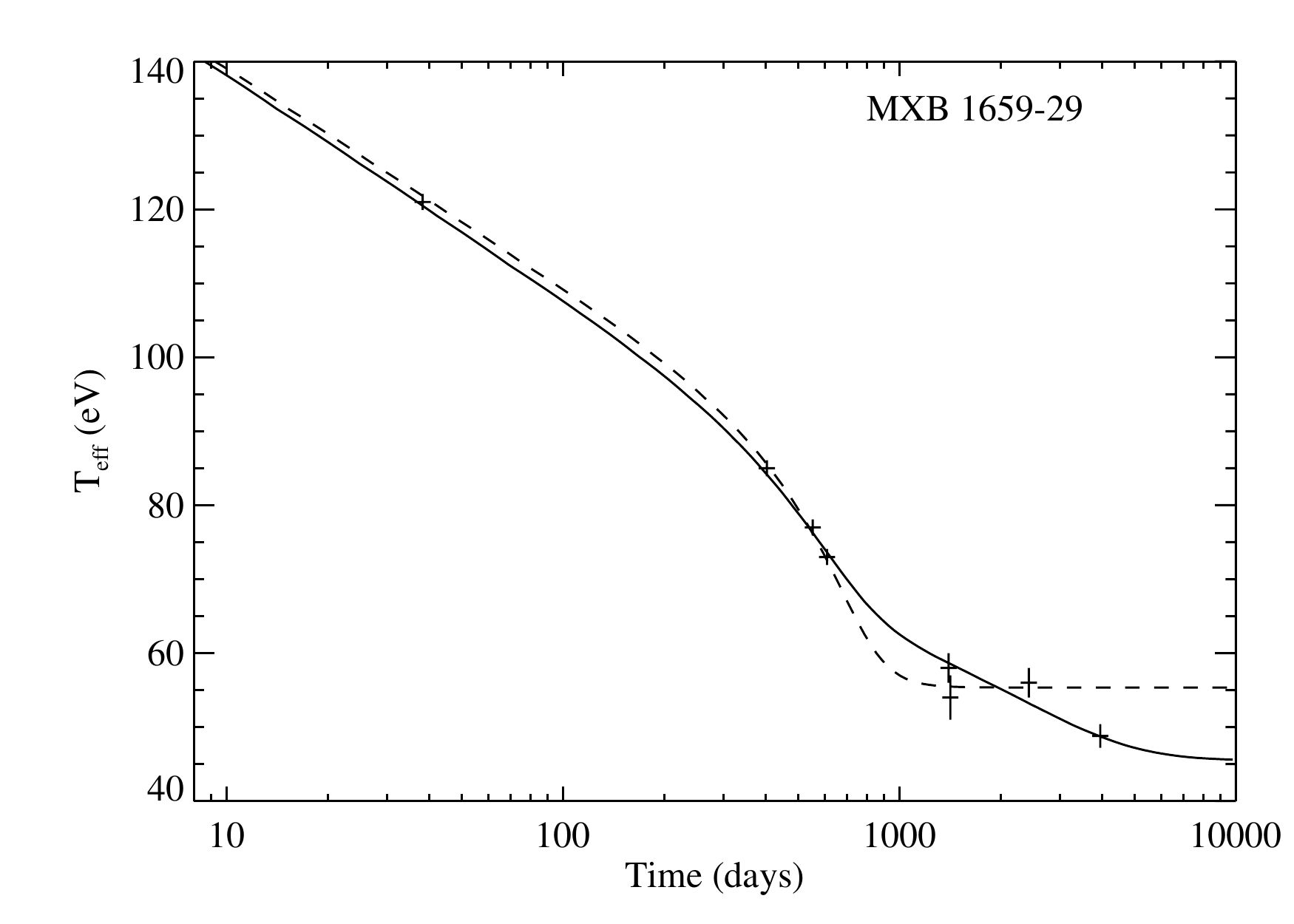}
\caption{Surface temperature $T_{\rm eff}$ vs time since accretion stopped for MXB 1659-29.  The dashed curve is the standard cooling model. It has $Q_{\rm imp}=3.5$ and core temperature $T_c = 3.05\times 10^7\ {\rm K}$. The solid curve has $Q_{\rm imp}=1.5$ for densities $<8\times 10^{13}\ {\rm g\ cm^{-3}}$ and $Q_{imp}$ = 30 for densities $>8\times 10^{13}\ {\rm g\ cm^{-3}}$. The core temperature in that model is lower, $T_c=2\times 10^7\ {\rm K}$.
\label{fig:tc}}
\end{figure}

Figure \ref{fig:tc} shows cooling curves for MXB~1659-29 with and without a low conductivity layer included. For the last temperature measurement, we use the value $49\pm 2\ {\rm eV}$ as in Figure~4 of Cackett et al.~\cite{Cackett2013}, which is likely an upper limit on the temperature, it could be significantly lower depending on the spectral model chosen (\cite{Cackett2013}. The model without a low conductivity layer is as shown in Figure 4 of  ref. \cite{Cackett2013} and is similar to the best fit of Brown \& Cumming \cite{Brown2009}. It has an impurity parameter in the inner crust of $Q_{\rm imp}=3.5$ and a core temperature $T_c=3.05\times 10^7\ {\rm K}$. The model with a low conductivity layer has $Q_{\rm imp}=30$ for $\rho>\rho_{\rm pasta}=8\times 10^{13}\ {\rm g\ cm^{-3}}$, leading to a dramatically longer cooling timescale for the inner crust, with the surface temperature continuing to drop for several thousand days. In order to bring the curve into agreement with the data, we have also lowered the impurity parameter for $\rho<\rho_{\rm pasta}$ to $Q_{\rm imp}=1.5$ and reduced the core temperature to $T_c=2\times 10^7\ {\rm K}$ to give a final surface temperature $\approx 45\ {\rm eV}$.   {\it The value $Q_{imp}=30$ for the pasta layer, that fits the Chandra observations, is in excellent agreement with our rough estimate $Q_{imp}\approx$ 40 in Eq. \ref{eq.qimp}.}

\begin{figure}
\centering
\includegraphics[width=0.5\textwidth]{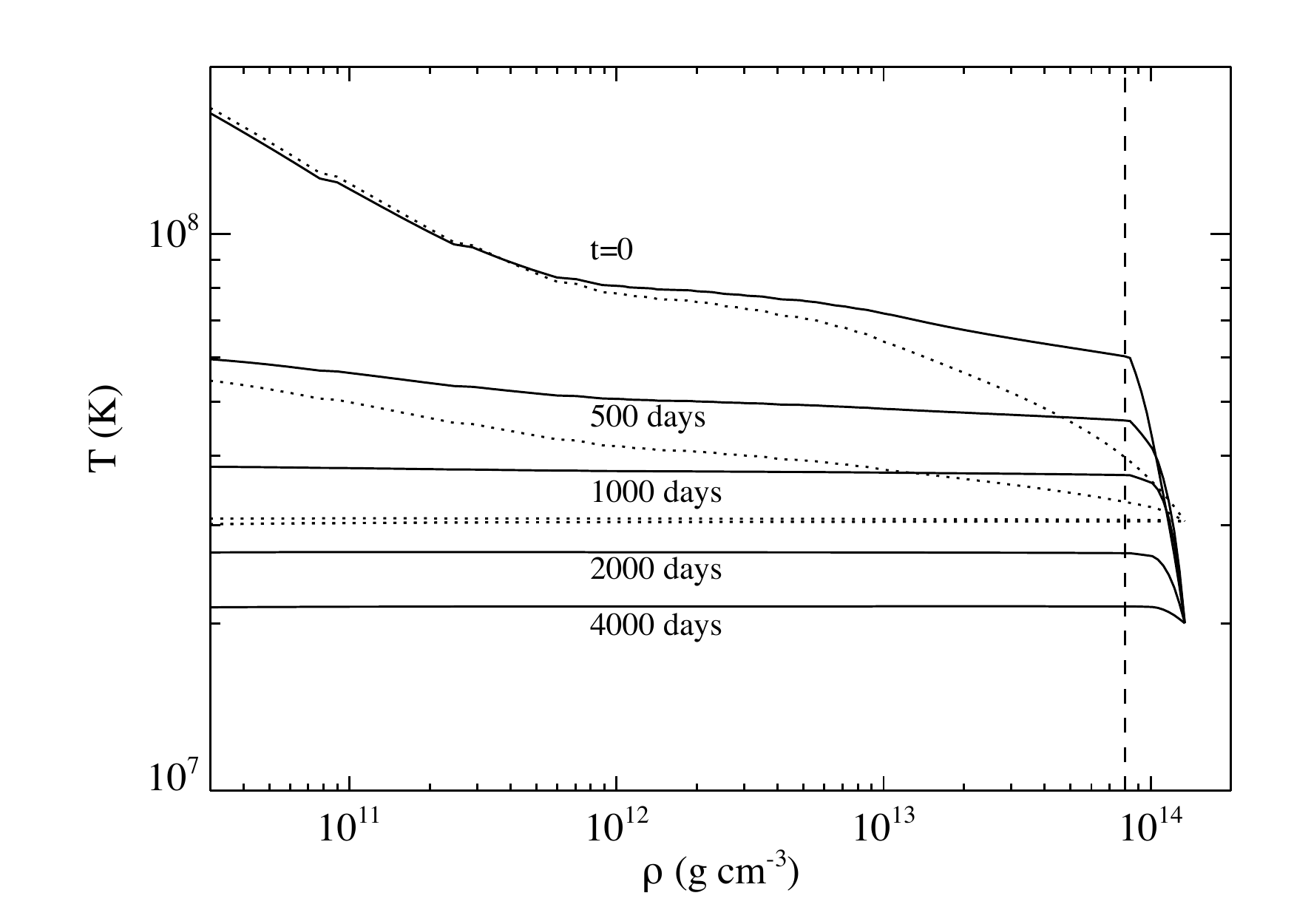}
\caption{Temperature profile as a function of time in quiescence for the models shown in Figure \ref{fig:tc} with (solid curves) and without (dotted curves) a low conductivity pasta layer. The vertical dashed line marks the outer boundary of the pasta layer $\rho_{\rm pasta}=8\times 10^{13}\ {\rm g\ cm^{-3}}$. After 1000 days, the model without a pasta layer has an isothermal crust (the crust has reached equilibrium with the core), whereas the model with a pasta layer continues to cool beyond 1000 days.
\label{fig:T10}}
\end{figure}

Figure \ref{fig:T10} shows the temperature profile of the crust at different times during the cooling phase.  Without a pasta layer, the crust is isothermal at the core temperature after approximately 1000 days; with a pasta layer, the crust is still isothermal for $\rho<\rho_{\rm pasta}$, but a significant temperature gradient remains across the pasta layer that slowly relaxes on timescales of thousands of days.   The thickness of the pasta layer is constrained by the requirement that the cooling curve drop fast enough to match the temperature decline at $\approx 300\ {\rm days}$. Brown \& Cumming \cite{Brown2009} assumed a constant $Q_{\rm imp}$ throughout the inner crust and found an upper limit $Q_{\rm imp}\lesssim 10$ after taking into account a range of surface gravities. This shows that the low conductivity layer cannot span the entire inner crust.  We find that to match the drop at a few hundred days and the late time drop in temperature at a few thousand days we need $\rho_{\rm pasta}\gtrsim 5\times 10^{13}\ {\rm g\ cm^{-3}}$. Note that this is for the particular choices $Q_{\rm imp}=30$ and surface gravity corresponding to a $1.6 M_\odot$, $11.2\ {\rm km}$ neutron star. 

Two other sources that have late time measurements are KS~1731-260 that was fit by Shternin et al. \cite{Shternin2007} and Brown \& Cumming \cite{Brown2009}, and XTE J1701-462 (Page \& Reddy \cite{Page2012}, \cite{Page2013}). However, we find that we can fit these sources equally well with or without a pasta layer. MXB~1659-29 is unique so far in having returned to equilibrium with the core in standard cooling models \cite{Cackett2008}, so that a drop in temperature requires an additional, low conductivity, component at the base of the crust.

Much remains to be done.  In future work we will calculate in more detail the effect of these defects on electron pasta scattering and determine if these defects do indeed significantly reduce the thermal and electrical conductivities.  If they do, one would like to know the number, distribution, and kind of topological defects that are likely to be present in nuclear pasta regions in neutron stars.  This may be difficult to determine, from first principles, because of the small energy differences.  Finally, one should study the dependence of defects on the interaction model, temperature, and proton fraction.     

Additional X-ray observations of late time crust cooling could help constrain the pasta thermal conductivity.  Although addressing this with accreting sources may be difficult, as it needs observations at thousands of days and the inner crust needs to be heated enough to get an observable temperature drop, meaning after a long duration transient.  One promising possibility is to look at magnetars. They show months long X-ray outbursts after which the flux slowly relaxes.  This has been modeled with crust cooling \cite{LET2002,PR2012,S2012,S2014}.
Most importantly, the outbursts have been seen to repeat on observable timescales. The magnetar SGR 1627-41 \cite{K2003,A2012}
showed a flux drop at approximately 2000 days after its 1998 outburst. Another, shorter, outburst occurred ten years later in 2008, and the flux decay is now being monitored. Whether it shows similar behavior at late times and the fact that there were two outbursts with different durations will constrain the allowed inner crust properties further.

In conclusion, large scale MD simulations find topological defects that form in nuclear pasta and are stable for long MD simulation times.  Electron scattering from these defects could possibly reduce both the electrical and thermal conductivity of nuclear pasta.  The thermal conductivity can be probed with crust cooling light curves that are sensitive to nuclear pasta properties for times from about three to ten years after accretion stops.   Indeed a late time Chandra observation of MXB 1659-29 is consistent with a low pasta thermal conductivity (although another interpretation is possible).     If additional crust cooling observations indeed support a low pasta thermal conductivity, this would suggest that the electrical conductivity might also be low.  This could lead to magnetic field decay in of order $10^6$ years, for field configurations supported by currents in the crust.

We thank the Institute for Nuclear Theory in Seattle, USA, and the International Space Science Institute in Bern, Switzerland for their hospitality.   Ed Brown and other members of the ISSI team  ``Probing Deep into the Neutron Star Crust with Transient Neutron-Star Low-Mass X-Ray Binaries" led by Dany Page are thanked for useful discussion as are Gerardo Ortiz and Herb Fertig.  This research was supported in part by DOE grants DE-FG02-87ER40365 (Indiana University) and DE-SC0008808 (NUCLEI SciDAC Collaboration).  AC is supported by an NSERC Discovery grant, is a member of the Centre de Recherche en Astrophysique du Qu\'ebec (CRAQ), and an Associate of the CIFAR Cosmology and Gravity program.


%

\end{document}